\documentclass[aps,pra,twocolumn,showpacs,superscriptaddress]{revtex4-1}

\usepackage{graphicx}
\usepackage{latexsym}
\usepackage{amsmath}
\usepackage{amssymb}

\begin{document}
\title{Entanglement in eight-qubit graph states}
\author{Ad\'{a}n Cabello}
\email{adan@us.es}
\affiliation{Departamento de F\'{\i}sica Aplicada II,
Universidad de Sevilla, E-41012 Sevilla, Spain}
\author{Antonio J. L\'{o}pez-Tarrida}
\affiliation{Departamento de F\'{\i}sica Aplicada II, Universidad de
Sevilla, E-41012 Sevilla, Spain}
\author{Pilar Moreno}
\affiliation{Departamento de F\'{\i}sica Aplicada II, Universidad de
Sevilla, E-41012 Sevilla, Spain}
\author{Jos\'{e} R. Portillo}
\affiliation{Departamento de Matem\'{a}tica Aplicada I, Universidad
de Sevilla, E-41012 Sevilla, Spain}
\date{\today}




\begin{abstract}
Any $8$-qubit graph state belongs to one of the $101$~equivalence
classes under local unitary operations within the Clifford group.
For each of these classes we obtain a representative which requires
the minimum number of controlled-$Z$ gates for its preparation, and
calculate the Schmidt measure for the $8$-partite split, and the
Schmidt ranks for all bipartite splits. This results into an
extension to $8$~qubits of the classification of graph states
proposed by Hein, Eisert, and Briegel [Phys. Rev.~A {\bf 69}, 062311
(2004)].
\end{abstract}


\maketitle


\section{Introduction}
\label{Sec1}


Graph states \cite{HEB04, HDERVB06} are a type of $n$-qubit pure
states that play several fundamental roles in quantum information
theory. In quantum error-correction, the stabilizer codes which
protect quantum systems from errors \cite{Gottesman96} can be
realized as graph codes \cite{SW02, Schlingemann02}. In
measurement-based (or one-way) quantum computation \cite{RB01},
graph states are the initial resources consumed during the
computation. Moreover, some graph states are universal resources for
quantum computation \cite{VMDB06}. In quantum simulation, graph
states allow us to demonstrate fractional braiding statistics of
anyons in an exactly solvable spin model \cite{HRD07}. Graph states
have been used in multipartite purification schemes \cite{DAB03}.
The Clifford group has been used for entanglement distillation
protocols \cite{DVDV03}. Graph states naturally lead to
Greenberger-Horne-Zeilinger (GHZ) or all-versus-nothing proofs of
Bell's theorem \cite{GHZ89, DP97, SASA05, GTHB05, Cabello05, CM07},
which can be converted into Bell inequalities which are maximally
violated by graph states \cite{Mermin90, Ardehali92, TGB06, Hsu06,
CGR07, GC08}. Some specific graph states are essential for several
quantum communication protocols, including entanglement-based
quantum key distribution \cite{Ekert91}, teleportation
\cite{BBCJPW93}, reduction of communication complexity \cite{CB97},
and secret sharing \cite{ZZHW98, MS08}.

In addition to all these applications, graph states also play a
fundamental role in the theory of entanglement. For $n \ge 4$
qubits, there is an infinite amount of different, inequivalent
classes of $n$-qubit pure entangled states. The graph state
formalism is a useful abstraction which permits a detailed (although
not exhaustive) classification of $n$-qubit entanglement of $n \ge
4$ qubits.

For all these reasons, a significant experimental effort is devoted
to the creation and testing of graph states of an increasing number
of qubits. On one hand, there are experiments of $n$-qubit
$n$-photon graph states up to $n=6$ \cite{WRRSWVAZ05, KSWGTUW05,
LZGGZYGYP07, LGGZCP07, PWSKPW07}. On the other hand, the combination
of two techniques, hyper-entanglement (i.e., entanglement in several
degrees of freedom, like polarization and linear momentum)
\cite{Kwiat97, KW98, CBPMD05, YZZYZZCP05, BLPK05, BDMVC06, VPMDB07}
and the sources of $4$, $5$, and $6$-photon entanglement using
parametric down-conversion \cite{WZ01, EGBKZW03, GBEKW03, BEGKCW04,
WARZ05, KSWGTUW05, GBKCW08} allows us to create $6$-qubit $4$-photon
graph states \cite{GLYXGGCPCP08, GYXGCLCP08}, $8$-qubit $4$-photon
graph states \cite{GLYXGGCPCP08}, and even $10$-qubit $5$-photon
graph states \cite{GLYXGGCPCP08}. The use of $4$-photon sources for
preparing $8$-qubit graph states is particularly suitable due to the
high visibility of the resulting states.

The classification and study of the entanglement properties of graph
states have been achieved, up to $7$ qubits, by Hein, Eisert, and
Briegel (HEB) in \cite{HEB04} (see also \cite{HDERVB06}). This
classification has been useful to identify new two-observer
all-versus-nothing proofs \cite{CM07}, new Bell inequalities
\cite{CGR07, GC08}, and has stimulated the preparation of several
graph states \cite{GLYXGGCPCP08}. The main purpose of this Letter is
to extend the classification in \cite{HEB04, HDERVB06} to $8$-qubit
graph states.

Up to $7$ qubits, there are $45$ classes of graph states that are
not equivalent under one-qubit unitary transformations. With $8$
qubits, there are $101$~new classes. All these classes have been
obtained by various researchers (see, e.g., \cite{Danielsen05}). The
purpose here is to classify them according to several relevant
physical properties for quantum information theory.

The Letter is organized as follows. In Sec.~\ref{Sec2} we define
qubit graph states and local complementation, which is the main
classifying tool. To establish an order between the equivalence
classes we will use the criteria proposed in \cite{HEB04, HDERVB06}.
These criteria are introduced in Sec.~\ref{Sec22}. In
Sec.~\ref{Sec3} we present our results. In Sec.~\ref{Sec4} we
present the conclusions and point out some pending problems.


\section{Basic concepts}
\label{Sec2}


\subsection{Graph state}


A $n$-qubit {\em graph state} $|G\rangle$ is a pure state associated
to a graph $G=(V, E)$ consisting of a set $V$ of $n$ vertices and a
set $E$ of edges connecting some of the vertices. Each vertex
represents a qubit. The graph $G$ provides both a recipe for
preparing $|G\rangle$ and a mathematical characterization of
$|G\rangle$.

The recipe for preparing the state is the following. First, prepare
each qubit in the state $|+\rangle =
\left(|0\rangle+|1\rangle\right)/\sqrt{2}$. Then, for each edge
connecting two qubits, $i$ and $j$, apply the controlled-$Z$ gate
between qubits $i$ and $j$, i.e., the unitary transformation
$C_Z=|00\rangle\langle00|+|01\rangle\langle01|+|10\rangle\langle10|-|11\rangle\langle11|$.

The mathematical characterization of $|G\rangle$ is the following.
The graph state $|G\rangle$ associated to the graph $G$ is the only
$n$-qubit state which fulfills
\begin{equation}
g_i |G\rangle= |G\rangle, \quad{\rm for}\quad i=1,\ldots,n,
\label{graphdef}
\end{equation}
where $g_i$ are the generators of the stabilizer group of the state,
defined as the set $\{s_j\}_{j=1}^{2^n}$ of all products of the
generators. Specifically, $g_i$ is the generator operator associated
to the vertex $i$, defined by
\begin{equation}
g_i:= X^{(i)}\bigotimes\nolimits_{j\in {\cal N}(i)} Z^{(j)},
\label{stabdef}
\end{equation}
where ${\cal N}(i)$ is the neighborhood of the vertex $i$, i.e.,
those vertices which are connected to $i$, and $X^{(i)}$ ($Z^{(i)}$)
denotes the Pauli matrix $\sigma_x$ ($\sigma_z$) acting on the $i$th
qubit.


\subsection{\label{localcomplementation}Local complementation}


For our purposes, the key point is that {\em local complementation}
(LC) is a simple transformation which leaves the entanglement
properties invariant.

Two $n$-qubit states, $|\phi\rangle$ and $|\psi\rangle$ have the
same $n$-partite entanglement if and only if there are $n$ one-qubit
unitary transformations $U_i$, such that $|\phi\rangle =
\bigotimes_{i=1}^n U_i |\psi\rangle$. If these one-qubit unitary
transformations belong to the Clifford group, then both states are
said to be local Clifford equivalent. The one-qubit Clifford group
is generated by the Hadamard gate
$H=\left(|0\rangle\langle0|+|0\rangle\langle1|+|1\rangle\langle0|-|1\rangle\langle1|\right)/\sqrt{2}$
and the phase gate $P=|0\rangle\langle0|+i |1\rangle\langle1|$.

Van den Nest, Dehaene, and De Moor found that the successive
application of a transformation with a simple graphical description
is sufficient to generate the complete equivalence class of graph
states under local unitary operations within the Clifford group
(hereafter simply referred as class or orbit) \cite{VDM04}. This
simple transformation is LC.

On the stabilizer, LC on the qubit $i$ induces the map $Y^{(i)}
\mapsto Z^{(i)}, Z^{(i)} \mapsto -Y^{(i)}$ on the qubit $i$, and the
map $X^{(j)} \mapsto -Y^{(j)}, Y^{(j)} \mapsto X^{(j)}$ on the
qubits $j\in {\cal N}(i)$ \cite{HDERVB06}. On the generators, LC on
the qubit $i$ maps the generators $g^{\rm{old}}_j$ with $j\in {\cal
N}(i)$ to $g^{\rm{new}}_j g^{\rm{new}}_{i}$.

Graphically, LC on the qubit $i$ acts as follows: One picks out the
vertex $i$ and inverts the neighborhood ${\cal N}(i)$ of $i$; i.e.,
vertices in the neighborhood which were connected become
disconnected and vice versa.

It has been shown by Van den Nest, Dehaene, and De Moor that for a
particular class of qubit graph states local unitary equivalence
implies local Clifford equivalence \cite{VDM05}. Moreover, numerical
results show that local Clifford equivalence coincides with local
unitary equivalence for qubit graph states associated with connected
graphs up to $n=7$ vertices \cite{HEB04, HDERVB06}. It should be
noted, however, that not all local unitary transformations between
graph states can be represented as successive LCs. A counterexample
with $n=27$ is described in \cite{JCWY07}.

Using LC, one can generate the orbits of all LC-inequivalent
$n$-qubit graph states. For a small $n$, the number of orbits has
been well known for a long time (see, e.g., \cite{Danielsen05}).
Specifically, for $n=8$, there are $101$~LC-inequivalent classes.


\section{\label{Sec22}Criteria for the classification}


Following HEB, the criteria for ordering the classes are: (a) number
of qubits, (b) minimum number of controlled-$Z$ gates needed for the
preparation, (c) the Schmidt measure, and (d) the rank indexes. For
instance, class No.~1 is the only one containing two-qubit graph
states, class No.~2 is the only one containing three-qubit graph
states \cite{HEB04, HDERVB06}. Classes No.~3 and No.~4 both have
$n=4$ qubits and require a minimum of $|E|$=3 controlled-$Z$ gates.
However, class No.~3 has Schmidt measure $E_S=1$, while class No.~4
has $E_S=2$.


\subsection{Minimum number of controlled-$Z$ gates for the preparation}


Different members of the same LC class require a different number of
controlled-$Z$ gates for their preparation starting from the state
$|+\rangle = \left(|0\rangle+|1\rangle\right)/\sqrt{2}$ for each
qubit. The first criterion for our classification is the minimum
number of controlled-$Z$ gates required for preparing one graph
state {\em within the LC class}. This corresponds to the number of
edges of the graph with the minimum number of edges within the LC
class, $|E|$. We will provide a representative with the minimum
number of edges for each LC class.


\subsection{Schmidt measure}


The {\em Schmidt measure} was introduced by Eisert and Briegel as a
tool for quantifying the genuine multipartite entanglement of
quantum systems \cite{EB01} (see also \cite{Severini06}). Any state
vector $|\psi\rangle \in \mathcal{H}^{(1)} \otimes \ldots \otimes
\mathcal{H}^{(N)}$ of a composite quantum system with $N$ components
can be represented as
\begin{equation}
|\psi\rangle=\sum_{i=1}^{R} \xi_{i} |\psi_{i}^{(1)}\rangle \otimes
\cdots \otimes |\psi_{i}^{(N)}\rangle, \label{SchDec}
\end{equation}
where $\xi_{i} \in \mathbb{C}$ for $i=1,\ldots,R$, and
$|\psi_{i}^{(j)}\rangle \in \mathcal{H}^{(j)}$, for $j=1,\ldots,N$.
The Schmidt measure associated with a state vector $|\psi\rangle$ is
then defined as
\begin{equation}
E_{S}(|\psi\rangle)=\log_{2}(r) \label{SchM},
\end{equation}
where $r$ is the minimal number $R$ of terms in the sum of
Eq.~(\ref{SchDec}) over all linear decompositions into product
states. In case of a two-component system ($N=2$), the minimal
number of product terms $r$ is given by the {\em Schmidt rank} of
the state $|\psi\rangle$. Hence, the Schmidt measure could be
considered a generalization of the Schmidt rank to multipartite
quantum systems [see Eq.~(\ref{Schrank}) below]. The Schmidt measure
can be extended to mixed states by means of a convex roof extension.
In this Letter, however, we will deal only with pure states.

Given a graph $G=(V,E)$, a {\em partition} of $V$ is any tuple
$(A_{1}, \ldots, A_{M})$ of disjoint subsets $A_{i}\subset V$, with
$\bigcup_{i=1}^{M} A_{i}=V$. In case $M=2$, we refer to the
partition as a {\em bipartition}, and denote it $(A,B)$. We will
write
\begin{equation}
(A_{1},\ldots,A_{N})\leq(B_{1},\ldots,B_{M}), \label{finerpart}
\end{equation}
if $(A_{1},\ldots,A_{N})$ is a {\em finer partition} than
$(B_{1},\ldots,B_{M})$, which means that every $A_{i}$ is contained
in some $B_{j}$. The latter is then a {\em coarser partition} than
the former. For any graph $G=(V,E)$, the partitioning where
$(A_{1},\ldots,A_{M})=V$ such that $|A_i|=1$, for every
$i=1,\ldots,M$, is referred to as the {\em finest partition}.

We must point out that $E_{S}$ is nonincreasing under a coarse
graining of the partitioning: If two components are merged in order
to form a new component, then the Schmidt measure can only decrease.
If we denote the Schmidt measure of a state vector $|\psi\rangle$
evaluated with respect to a partitioning $(A_{1},\ldots,A_{N})$ as
$E_{S}^{(A_{1},\ldots,A_{N})}(|\psi\rangle)$, meaning that the
respective Hilbert spaces are those of the grains of the
partitioning, then the nonincreasing property of $E_{S}$ can be
expressed as
\begin{equation}
E_{S}^{(A_{1},\ldots,A_{N})}(|\psi\rangle)\geq
E_{S}^{(B_{1},\ldots,B_{M})}(|\psi\rangle), \label{nonincreasing}
\end{equation}
if $(A_{1},\ldots,A_{N})\leq (B_{1},\ldots,B_{M})$.

Let $(A,B)$ be a bipartition (i.e., $A \cup B=V; A \cap
B=\emptyset$) of a graph $G=(V,E)$, with $V=\{1,\ldots,N\}$, and let
us denote the adjacency matrix of the graph by $\Gamma$, i.e., the
symmetric matrix with elements
\begin{equation}
\Gamma_{ij}=
\begin{cases}
1, &\text{if $(i,j)\in E$,}\\
0, &\text{otherwise.}
\end{cases}\label{Gamma}
\end{equation}
When we are dealing with a bipartition, it is useful to label the
vertices of the graph so that $A=\{1,\ldots,p\}$,
$B=\{p+1,\ldots,N\}$. Then, we can decompose the adjacency matrix
$\Gamma$ into submatrices $\Gamma_{A}$, $\Gamma_{B}$ (that represent
edges within $A$ and edges within $B$), and $\Gamma_{AB}$ (the
$|A|\times|B|$ off-diagonal submatrix of the adjacency matrix
$\Gamma$ that represents those edges between $A$ and $B$),
\begin{eqnarray}
\left(
\begin{matrix}
\Gamma_{A} & \Gamma_{AB} \\
\Gamma_{AB}^{T} & \Gamma_{B}
\end{matrix}
\right) & = & \Gamma. \label{Gmatrix}
\end{eqnarray}
The Schmidt rank $SR_{A}(G)$ of a graph state $|G\rangle$
represented by the graph $G=(V,E)$, with respect to the bipartition
$(A,B)$, is given by the binary rank [i.e., the rank over $GF(2)$]
of the submatrix $\Gamma_{AB}$,
\begin{equation}
SR_{A}(G)={\rm rank}_{\mathcal{F}_{2}}(\Gamma_{AB}). \label{Schrank}
\end{equation}
It follows straightforwardly from the definition that
$SR_{A}(G)=SR_{B}(G)$, because the different bipartitions are fixed
by choosing the smaller part, say $A$, of the bipartition $(A,B)$,
which gives $2^{N-1}$ bipartitions.


\subsection{Rank indexes}


While calculating the Schmidt rank with respect to all possible
bipartitions of a given graph, let us count how many times a certain
rank occurs in all the bipartite splits, and then classify this
information according to the number of vertices in $A$, the smaller
part of the split under consideration. There is a compact way to
express this information, the so-called {\em rank indexes}
\cite{HEB04, HDERVB06}. The rank index for all the bipartite splits
with $p$ vertices in the smaller part $A$ is given by the $p$-tuple
\begin{equation}
RI_{p}=(\nu_{p}^{p},\ldots,\nu_{1}^{p})=[\nu_{j}^{p}]_{j=p}^{1},
\label{rankindex}
\end{equation}
where $\nu_{j}^{p}$ is the number of times in which $SR_{A}(G)=j$,
with $|A|=p$, occurs.


\section{Procedures and results}
\label{Sec3}


The main results of the Letter are summarized in Fig.~\ref{Fig1} and
Table~\ref{MainTable}. In the following, we provide details on the
calculations leading to these results.


\subsection{Orbits under local complementation}


We have generated all LC orbits for $n=8$ and calculated the number
of non-isomorphic graphs in each LC orbit, denoted by $|LC|$. These
numbers are counted up to isomorphism.

In addition, for each orbit, we have calculated a representative
with the minimum number of edges $|E|$. As representative, we have
chosen the one (or one of those) with the minimum number of edges
and the minimum maximum degree (i.e., number of edges incident with
a vertex). This means that the graph state associated to this graph
requires the minimum number of controlled-$Z$ gates for its
preparation, and the minimum preparation depth (i.e., its
preparation requires a minimum number of steps) \cite{MP04}. All the
representatives of each of the $101$~orbits are illustrated in
Fig.~\ref{Fig1}. $|LC|$ and $|E|$ are in Table~\ref{MainTable}.


\subsection{Bounds to the Schmidt measure}


It is a well-known fact that for any measure of multiparticle
entanglement proposed so far, including the Schmidt measure $E_{S}$,
the computation is exceedingly difficult for general states. In
order to determine $E_{S}$, one has to show that a given
decomposition in Eq.~(\ref{SchDec}) with $R$ terms is minimal. For a
general state, the minimization problem involved can be a very
difficult problem of numerical analysis, which scales exponentially
in the number of parties $N$ as well as in the degree of
entanglement of the state itself. Nevertheless, this task becomes
feasible if we restrict our attention to graph states. HEB
established several upper and lower bounds for the Schmidt measure
in graph theoretical terms \cite{HEB04, HDERVB06}. These bounds make
possible to determine the Schmidt measure for a large number of
graphs of practical importance, because in many cases the bounds
proposed are easily computable and, remarkably, the upper and lower
bounds frequently coincide.


\subsubsection{Pauli persistency and size of the minimal vertex cover}


For any graph state $|G\rangle$, upper bounds for its Schmidt
measure $E_{S}(|G\rangle)$ are the {\em Pauli persistency} $PP(G)$
and the {\em size of the minimal vertex cover} $VC(G)$,
\begin{equation}
E_{S}(|G\rangle) \le PP(G) \le VC(G). \label{bounds}
\end{equation}
The Pauli persistency is the minimal number of local Pauli
measurements necessary to disentangle a graph state. Concerning this
question, HEB described graphical transformation rules when local
Pauli measurements are applied \cite{HEB04, HDERVB06}.

\begin{figure}[htb]
\centerline{\includegraphics[width=0.75
\columnwidth]{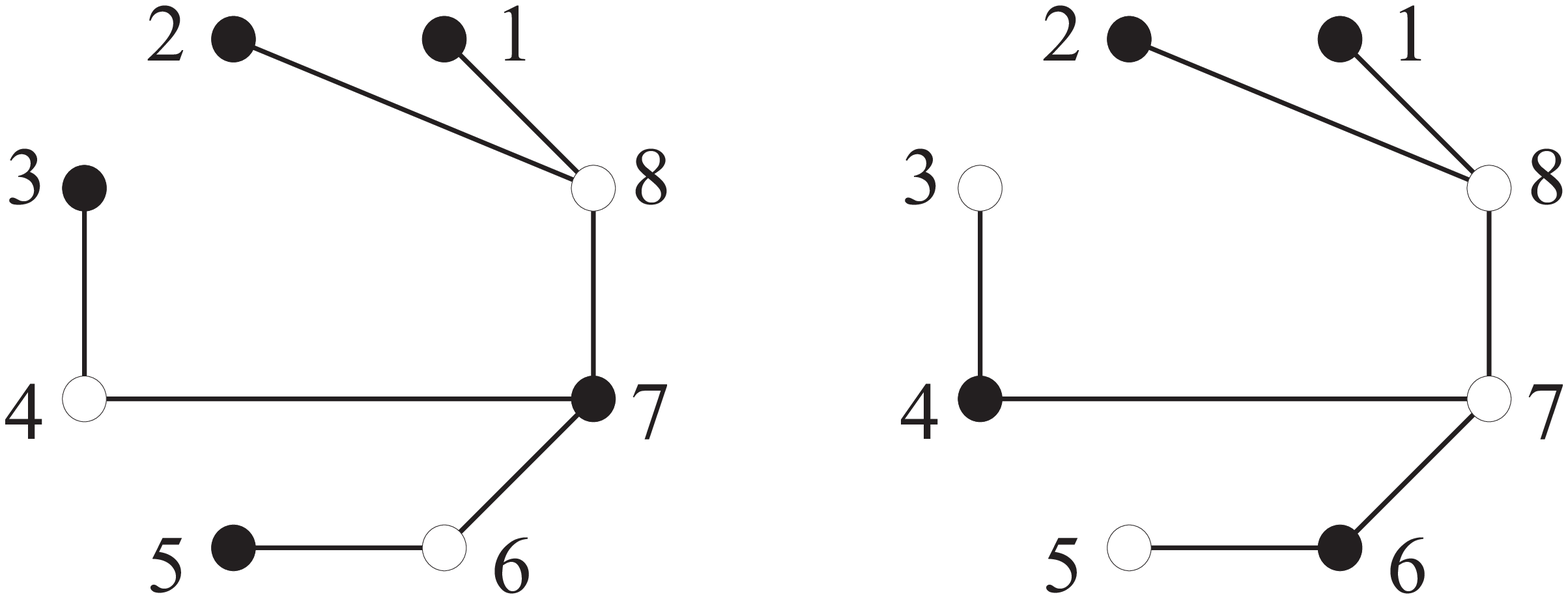}} \caption{\label{Fig0} The set
of vertices $4$, $6$, and $8$ is the minimal vertex cover of the
graph (left). The set of vertices $3$, $5$, $7$, and $8$, is a
vertex cover of the graph, but is not minimal, it has size $4$
(right).}
\end{figure}

A vertex cover is a concept from graph theory: It is any subset $V'
\subseteq V$ of vertices in a graph $G$ to which {\em any} edge of
$G$ is incident (see Fig. \ref{Fig0}). Therefore, the minimal vertex
cover of a graph is the smallest one, whose size is denoted by
$VC(G)$. According to the graphical rules for the Pauli
measurements, since each $\sigma_z$ measurement simply deletes all
edges incident to a vertex, the size of the minimal vertex cover
would equal the Pauli persistency, provided that we restrict the
Pauli measurements to $\sigma_z$ measurements. Nevertheless, in
graphs with many edges, i.e., very connected, a proper combination
of $\sigma_x$, $\sigma_y$, and $\sigma_z$ measurements could provide
a more efficient disentangling sequence, giving a better upper bound
$PP(G)$ for the Schmidt measure. See \cite{HEB04, HDERVB06} for
details.


\subsubsection{Maximal Schmidt rank}


For any graph state $|G\rangle$, a lower bound for the Schmidt
measure $E_{S}(|G\rangle)$ is the {\em maximal Schmidt rank},
\begin{equation}
SR_{\rm max}(G) \le E_{S}(|G\rangle). \label{bounds2}
\end{equation}
While calculating the Schmidt rank with respect to all possible
bipartitions of a given graph $G=(V,E)$, if one maximizes the
Schmidt rank over all bipartitions $(A,B)$ of the graph, and takes
into account the nonincreasing property of $E_{S}|(G)\rangle$ [see
Eq.~(\ref{nonincreasing})], then one obtains a lower bound for the
Schmidt measure with respect to the finest partitioning. This lower
bound is the maximal Schmidt rank,
\begin{equation}
SR_{\rm max}(G):= \max_{A\subseteq V} SR_{A}(G).\label{Schrankmax}
\end{equation}
According to the definition of Schmidt rank, the maximal Schmidt
rank for any state is, at most, $\lfloor\frac{N}{2}\rfloor$, i.e.,
the largest integer less than or equal to $\frac{N}{2}$.


\subsubsection{Addition and deletion of edges and
vertices}


As we mentioned in Sec.~\ref{localcomplementation}, applying the
LC-rule does not change the Schmidt measure $E_{S}$. It is
interesting to remark that other local changes to the graph, such as
the deletion of edges or vertices, have only a limited effect on
$E_{S}$. This fact is established by HEB \cite{HEB04} in what they
call the {\em edge/vertex rule}: On one hand, by deleting (or
adding) an edge $e$ between two vertices of a graph $G$ the Schmidt
measure of the resulting graph $G'=G\pm{e}$ can at most decrease (or
increase) by 1. On the other hand, if a vertex $v$ (including all
its incident edges) is deleted, the Schmidt measure of the resulting
graph $G'=G-{v}$ cannot increase, and will at most decrease by one.
If $E_S (|G+e\rangle)$ denotes the Schmidt measure of the graph
state corresponding to the graph $G+e$, then the previous rules can
be summarized as
\begin{subequations}
\begin{align}
& E_S (|G+e\rangle) \le E_S (|G\rangle)+1, \\
& E_S (|G-e\rangle) \ge E_S (|G\rangle)-1, \\
& E_S (|G-v\rangle) \ge E_S (|G\rangle)-1.
\end{align}
\end{subequations}
We have used these rules in two ways: Firstly, as an internal test
to check our calculations, comparing pairs of graphs connected by a
sequence of addition or deletion of edges/vertices; and secondly, as
a useful tool that, in some graphs, has enabled us to go from a
bounded to a determined value for the Schmidt measure, once again by
comparison between a problematic graph $G$ and a resulting graph
$G'$ (typically obtained by edge or vertex deletion) of a known
Schmidt measure.


\subsubsection{Schmidt measure in some special types of graphs}


There are some special types of graph states in which lower and
upper bounds for the Schmidt measure coincide (see \cite{HEB04}),
giving directly a determined value for $E_{S}$. Since the maximal
Schmidt rank for any state can be at most
$\lfloor\frac{N}{2}\rfloor$, and restricting ourselves to states
with coincident upper and lower bounds to $E_{S}$, it is true that
$SR_{\rm
max}(G)=E_{S}(|G\rangle)=PP(G)=VC(G)\leq\lfloor\frac{N}{2}\rfloor$.
This is the case for GHZ states, and states represented by trees,
rings with an even number of vertices, and clusters. In our work we
have used the following results concerning GHZ states and trees:

(a) The Schmidt measure for any multipartite GHZ state is 1.

(b) A {\em tree} $T$ is a graph that has no cycles. The Schmidt
measure of the corresponding graph state $|T\rangle$ is the size of
its minimal vertex cover: $E_{S}(|T\rangle)=VC(T)$.

There is another interesting kind of graphs for our purposes, the
so-called {\em 2-colorable graphs}. A graph is said to be
2-colorable when it is possible to perform a {\em proper 2-coloring}
on it: This is a labeling $V\rightarrow\{1,2\}$, such that all
connected vertices are associated with a different element from
$\{1,2\}$, which can be identified with two colors. It is a
well-known fact in graph theory that a necessary and sufficient
criterion for a graph to be 2-colorable is that it does not contain
any cycles of odd length. Mathematicians call these graphs {\em
bipartite graphs} due to the fact that the whole set of vertices can
be distributed into two disjoint subsets $A$ and $B$, such that no
two vertices within the same subset are connected, and therefore
every edge connects a vertex in $A$ with a vertex in $B$.

HEB \cite{HEB04} provided lower and upper bounds for the Schmidt
measure that could be applied to graph states represented by
2-colorable graphs:
\begin{equation}
\frac{1}{2} {\rm rank}_{\mathcal{F}_{2}}(\Gamma)\leq
E_{S}(|G\rangle)\leq \lfloor\frac{|V|}{2}\rfloor,
\label{bounds2-col}
\end{equation}
where $\Gamma$ is the adjacency matrix of the 2-colorable graph. If
$\Gamma$ is invertible, then
\begin{equation}
E_{S}(|G\rangle)= \lfloor\frac{|V|}{2}\rfloor.
\end{equation}
Besides, HEB pointed out that any graph $G$ which is not 2-colorable
can be turned into a 2-colorable one $G'$ by simply deleting the
appropriate vertices on cycles with odd length present in $G$. The
identification of this graphical action with the effect of a
$\sigma_{z}$ measurement on qubits corresponding to such vertices
yields new upper bounds for $E_{S}(|G\rangle)$:
\begin{equation}
E_{S}(|G\rangle)\leq E_{S}(|G'\rangle)+M \leq
\lfloor\frac{|V-M|}{2}\rfloor+M \leq \lfloor\frac{|V|+M}{2}\rfloor,
\label{boundsnot2-col}
\end{equation}
where $M$ is the number of removed vertices. We have used these new
bounds in some graphs as a tool to check our calculations.


\begin{figure*}[htb]
\centerline{\includegraphics[width=1.60
\columnwidth]{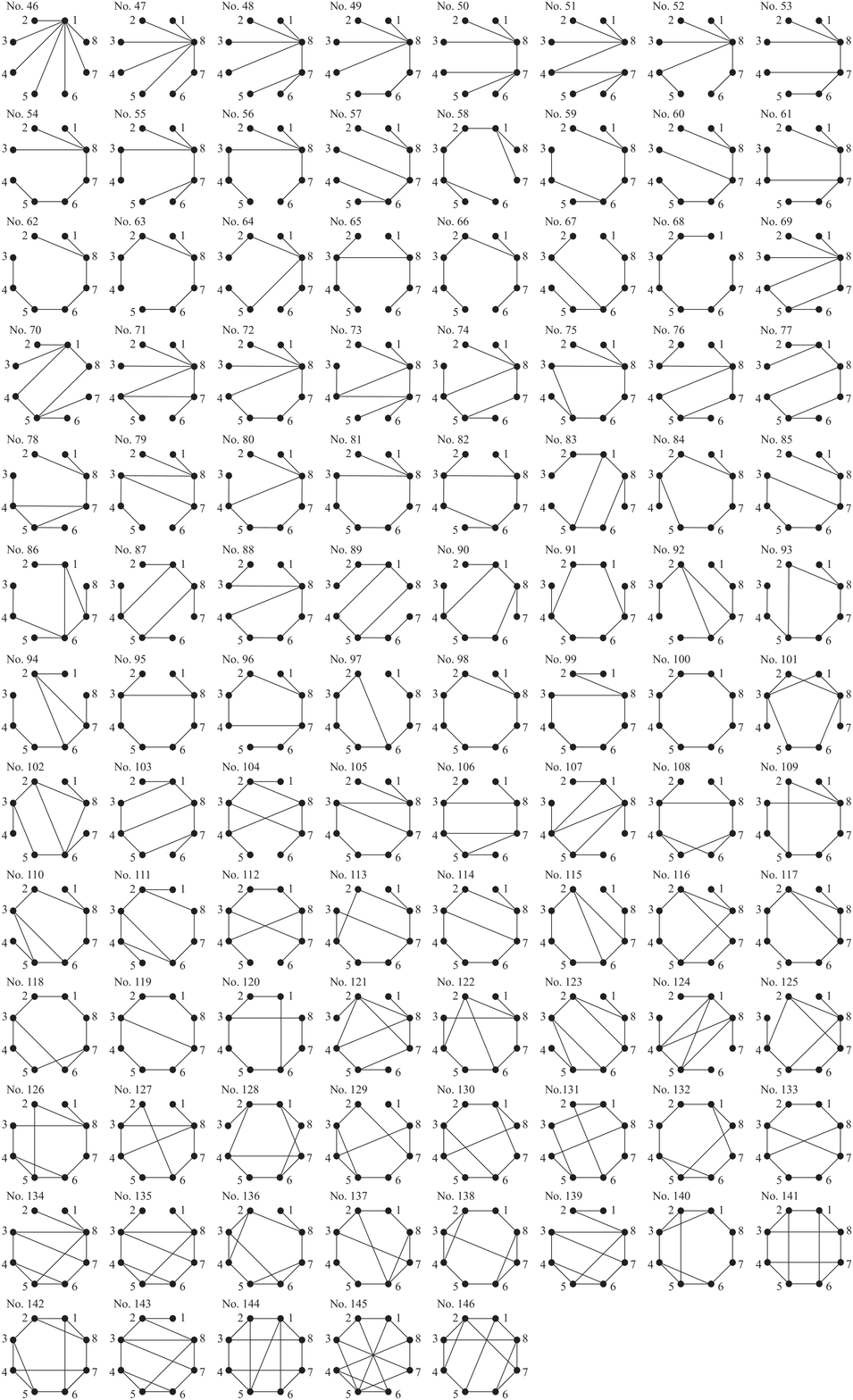}} \caption{\label{Fig1}
Graphs associated to the $101$~classes on $8$-qubit graph
states inequivalent under local complementation and graph
isomorphism. We have chosen as representative of the class the
one (or one of those) with minimum number of edges and minimum
maximum degree (i.e., number of edges incident with a vertex),
which means that it requires the minimum number of
controlled-$Z$ gates in the preparation and minimum preparation
depth.}
\end{figure*}


\begin{table*}[htb]
\caption{\label{MainTable}Entanglement of the $101$~classes of
$8$-qubit graph states. No. is the number of the class; it is
assigned attending to $|E|$, $E_S$, and $RI_j$; the numbering starts
at the point in which the one in Refs.~\cite{HEB04, HDERVB06} ends.
$|LC|$ is the number of nonisomorphic elements of the class. $|E|$
is the number of edges of those representatives with the minimum
number of edges. $E_S$ is the Schmidt measure (or its lower and
upper bounds). $RI_j$ is the rank index with $j$ qubits in the
smaller bipartition (i.e., the number of bipartite splits in which a
certain rank occurs; ranks appear in decreasing order from left to
right).
2-col indicates whether
or not a $2$-colorable representative exists.}
\begin{ruledtabular}
{\begin{tabular}{cccccccc|cccccccc} No. & $|LC|$ & $|E|$ & $E_S$ &
$RI_4$ & $RI_3$ & $RI_2$ & 2-col$\;$ & $\;$No. & $|LC|$ & $|E|$ &
$E_S$ & $RI_4$ & $RI_3$ & $RI_2$ &
2-col\\
\hline \hline
46 & 2 & 7 & 1 & (0,0,0,35) & (0,0,56) & (0,28) & yes       &  97 & 154 & 8 & 4 & (8,22,4,1) & (42,13,1) & (27,1) & no\\
47 & 6 & 7 & 2 & (0,0,20,15) & (0,30,26) & (12,16) & yes    &  98 & 542 & 8 & 4 & (8,22,5,0) & (42,14,0) & (27,1) & no\\
48 & 6 & 7 & 2 & (0,0,30,5) & (0,45,11) & (15,13) & yes     &  99 & 300 & 8 & 4 & (12,16,7,0) & (44,11,1) & (27,1) & yes\\
49 & 16 & 7 & 2 & (0,0,30,5) & (0,45,11) & (17,11) & yes    & 100 & 214 & 8 & 4 & (14,17,4,0) & (48,8,0) & (28,0) & yes\\
50 & 4 & 7 & 2 & (0,0,34,1) & (0,48,8) & (16,12) & yes      & 101 & 14 & 9 & 3 & (0,20,15,0) & (24,32,0) & (25,3) & no\\
51 & 16 & 7 & 2 & (0,0,34,1) & (0,51,5) & (19,9) & yes      & 102 & 66 & 9 & 3 & (0,28,7,0) & (32,24,0) & (25,3) & no\\
52 & 10 & 7 & 3 & (0,12,16,7) & (16,28,12) & (20,8) & yes   & 103 & 66 & 9 & 3 & (0,30,5,0) & (36,20,0) & (26,2) & yes\\
53 & 16 & 7 & 3 & (0,12,22,1) & (16,34,6) & (20,8) & yes    & 104 & 6 & 9 & 3 & (0,32,0,3) & (32,24,0) & (24,4) & yes\\
54 & 44 & 7 & 3 & (0,12,22,1) & (16,35,5) & (21,7) & yes    & 105 & 57 & 9 & $3<4$ & (0,30,5,0) & (36,19,1) & (25,3) & no\\
55 & 16 & 7 & 3 & (0,18,14,3) & (18,33,5) & (21,7) & yes    & 106 & 28 & 9 & 4 & (8,18,9,0) & (38,18,0) & (25,3) & no \\
56 & 44 & 7 & 3 & (0,18,14,3) & (22,29,5) & (23,5) & yes    & 107 & 17 & 9 & 4 & (8,20,6,1) & (32,24,0) & (24,4) & no\\
57 & 10 & 7 & 3 & (0,18,15,2) & (18,36,2) & (21,7) & yes    & 108 & 72 & 9 & 4 & (8,20,7,0) & (36,20,0) & (25,3) & no\\
58 & 25 & 7 & 3 & (0,18,16,1) & (18,36,2) & (22,6) & yes    & 109 & 87 & 9 & 4 & (8,20,7,0) & (40,16,0) & (27,1) & no\\
59 & 44 & 7 & 3 & (0,18,16,1) & (22,31,3) & (23,5) & yes    & 110 & 114 & 9 & 4 & (8,22,5,0) & (40,16,0) & (26,2) & no\\
60 & 44 & 7 & 3 & (0,24,9,2) & (24,30,2) & (23,5) & yes     & 111 & 372 & 9 & 4 & (8,22,5,0) & (40,16,0) & (26,2) & no\\
61 & 26 & 7 & 3 & (0,24,10,1) & (28,25,3) & (23,5) & yes    & 112 & 70 & 9 & 4 & (8,24,2,1) & (40,16,0) & (26,2) & no\\
62 & 120 & 7 & 3 & (0,24,10,1) & (28,26,2) & (24,4) & yes   & 113 & 264 & 9 & 4 & (8,24,3,0) & (44,12,0) & (27,1) & no\\
63 & 66 & 7 & 3 & (0,26,7,2) & (30,24,2) & (25,3) & yes     & 114 & 542 & 9 & 4 & (8,24,3,0) & (44,12,0) & (27,1) & no\\
64 & 14 & 7 & 4 & (8,12,12,3) & (32,18,6) & (24,4) & yes    & 115 & 156 & 9 & 4 & (12,18,5,0) & (46,9,1) & (27,1) & no\\
65 & 25 & 7 & 4 & (8,12,14,1) & (32,20,4) & (24,4) & yes    & 116 & 174 & 9 & 4 & (12,20,3,0) & (46,10,0) & (27,1) & no\\
66 & 120 & 7 & 4 & (8,14,12,1) & (34,19,3) & (25,3) & yes   & 117 & 542 & 9 & 4 & (12,20,3,0) & (46,10,0) & (27,1) & no\\
67 & 72 & 7 & 4 & (8,16,10,1) & (36,17,3) & (25,3) & yes    & 118 & 262 & 9 & 4 & (12,20,3,0) & (48,8,0) & (28,0) & no\\
68 & 172 & 7 & 4 & (8,18,8,1) & (38,16,2) & (26,2) & yes    & 119 & 802 & 9 & 4 & (14,19,2,0) & (50,6,0) & (28,0) & no\\
69 & 10 & 8 & 2 & (0,0,34,1) & (0,52,4) & (20,8) & yes      & 120 & 117 & 9 & 4 & (16,16,3,0) & (50,6,0) & (28,0) & yes\\
70 & 10 & 8 & 2 & (0,0,35,0) & (0,54,2) & (21,7) & yes      & 121 & 10 & 10 & 3 & (0,32,2,1) & (32,24,0) & (24,4) & no\\
71 & 10 & 8 & 3 & (0,12,22,1) & (16,36,4) & (20,8) & no     & 122 & 37 & 10 & 3 & (0,32,3,0) & (40,16,0) & (27,1) & yes\\
72 & 21 & 8 & 3 & (0,12,22,1) & (16,36,4) & (22,6) & no     & 123 & 36 & 10 & 4 & (8,22,5,0) & (44,12,0) & (26,2) & no\\
73 & 10 & 8 & 3 & (0,18,17,0) & (18,36,2) & (21,7) & no     & 124 & 7 & 10 & 4 & (8,24,0,3) & (32,24,0) & (24,4) & no\\
74 & 44 & 8 & 3 & (0,18,17,0) & (22,32,2) & (23,5) & yes    & 125 & 103 & 10 & 4 & (8,24,3,0) & (42,14,0) & (26,2) & no\\
75 & 66 & 8 & 3 & (0,18,17,0) & (22,33,1) & (24,4) & no     & 126 & 46 & 10 & 4 & (8,24,3,0) & (44,12,0) & (27,1) & no\\
76 & 26 & 8 & 3 & (0,20,14,1) & (24,30,2) & (24,4) & yes    & 127 & 170 & 10 & 4 & (8,26,1,0) & (46,10,0) & (27,1) & no\\
77 & 26 & 8 & 3 & (0,24,10,1) & (24,31,1) & (23,5) & yes    & 128 & 74 & 10 & 4 & (12,20,3,0) & (46,10,0) & (27,1) & yes\\
78 & 28 & 8 & 3 & (0,24,10,1) & (28,27,1) & (23,5) & no     & 129 & 340 & 10 & 4 & (12,22,1,0) & (48,8,0) & (27,1) & no\\
79 & 44 & 8 & 3 & (0,24,11,0) & (28,26,2) & (23,5) & no     & 130 & 254 & 10 & 4 & (12,22,1,0) & (50,6,0) & (28,0) & no\\
80 & 132 & 8 & 3 & (0,24,11,0) & (28,27,1) & (24,4) & no    & 131 & 433 & 10 & 4 & (14,21,0,0) & (52,4,0) & (28,0) & no\\
81 & 114 & 8 & 3 & (0,24,11,0) & (30,25,1) & (25,3) & yes   & 132 & 476 & 10 & 4 & (16,18,1,0) & (52,4,0) & (28,0) & no\\
82 & 72 & 8 & 3 & (0,26,9,0) & (30,26,0) & (25,3) & no      & 133 & 28 & 10 & $4<5$ & (12,22,0,1) & (48,8,0) & (28,0) & no\\
83 & 72 & 8 & 3 & (0,28,6,1) & (32,23,1) & (25,3) & yes     & 134 & 9 & 11 & $3<4$ & (0,30,5,0) & (40,15,1) & (25,3) & no\\
84 & 198 & 8 & 3 & (0,28,7,0) & (34,22,0) & (26,2) & yes    & 135 & 39 & 11 & 4 & (8,26,1,0) & (44,12,0) & (26,2) & no\\
85 & 56 & 8 & $3<4$ & (0,30,4,1) & (34,21,1) & (25,3) & no  & 136 & 46 & 11 & 4 & (12,20,3,0) & (50,6,0) & (27,1) & no\\
86 & 28 & 8 & 4 & (8,16,10,1) & (32,22,2) & (24,4) & no     & 137 & 208 & 11 & $4<5$ & (16,18,1,0) & (52,4,0) & (28,0) & no\\
87 & 10 & 8 & 4 & (8,16,10,1) & (32,24,0) & (24,4) & yes    & 138 & 298 & 11 & $4<5$ & (18,17,0,0) & (54,2,0) & (28,0) & no\\
88 & 56 & 8 & 4 & (8,16,10,1) & (36,18,2) & (26,2) & no     & 139 & 24 & 11 & $4<5$ & (20,10,5,0) & (50,5,1) & (27,1) & no\\
89 & 66 & 8 & 4 & (8,16,11,0) & (36,18,2) & (25,3) & yes    & 140 & 267 & 11 & $4<5$ & (20,14,1,0) & (54,2,0) & (28,0) & no\\
90 & 72 & 8 & 4 & (8,18,9,0) & (34,22,0) & (25,3) & no      & 141 & 4 & 12 & 4 & (28,0,7,0) & (56,0,0) & (28,0) & no\\
91 & 63 & 8 & 4 & (8,18,9,0) & (36,20,0) & (26,2) & yes     & 142 & 22 & 12 & $4<5$ & (14,21,0,0) & (56,0,0) & (28,0) & no\\
92 & 66 & 8 & 4 & (8,18,9,0) & (38,16,2) & (25,3) & no      & 143 & 46 & 12 & $4<5$ & (20,12,3,0) & (50,6,0) & (27,1) & yes\\
93 & 176 & 8 & 4 & (8,18,9,0) & (38,17,1) & (26,2) & no     & 144 & 28 & 13 & 4 & (28,4,3,0) & (56,0,0) & (28,0) & no\\
94 & 76 & 8 & 4 & (8,20,6,1) & (36,19,1) & (25,3) & no      & 145 & 7 & 13 & $4<5$ & (16,18,1,0) & (56,0,0) & (28,0) & no\\
95 & 194 & 8 & 4 & (8,20,7,0) & (38,18,0) & (26,2) & yes    & 146 & 51 & 13 & $4<5$ & (24,10,1,0) & (56,0,0) & (28,0) & no\\
96 & 352 & 8 & 4 & (8,20,7,0) & (40,15,1) & (26,2) & no    &
\end{tabular}}
\end{ruledtabular}
\end{table*}


\section{\label{Sec4}Conclusions, open problems, and future developments}


To sum it all up, we have extended to $8$ qubits the classification
of the entanglement of graph states proposed in \cite{HEB04} for
$n<8$ qubits. Notice that for $n=8$ we have $101$ classes, while for
$n<8$ there are only 45 classes. For each of these classes we obtain
a representative which requires the minimum controlled-$Z$ gates for
its preparation (see Fig.~\ref{Fig1}), and calculate the Schmidt
measure for the $8$-partite split (which measures the genuine
$8$-party entanglement of the class), and the Schmidt ranks for all
bipartite splits (see Table~\ref{MainTable}).

This classification will help us to obtain new all-versus-nothing
proofs of Bell's theorem \cite{CM07} and new Bell inequalities.
Specifically, any $8$-qubit graph state belonging to a class with a
representative with $7$ edges (i.e., a tree) has a specific type of
Bell inequality \cite{GC08}. More generally, it will help us to
investigate the nonlocality (i.e., the non-simulability of the
predictions of quantum mechanics by means of non-local hidden
variable models) of graph states \cite{CGR07}.

Extending the classification in \cite{HEB04} a further step sheds
some light on the limitations of the method of classification. The
criteria used in \cite{HEB04} to order the classes (see
Sec.~\ref{Sec22}) already failed to distinguish all classes in
$n=7$. For instance, classes No.~40, No.~42, and No.~43 in
\cite{HEB04, HDERVB06} have the same number of qubits, require the
same minimum number of controlled-$Z$ gates for the preparation, and
have the same Schmidt measure and rank indexes. The same problem
occurs between classes No.~110 and No.~111, between classes No.~113
and No.~114, and between classes No.~116 and No.~117 in our
classification (see Table~\ref{MainTable}). Following \cite{HEB04,
HDERVB06}, we have placed the class with lower $|LC|$ in the first
place. However, this solution is not satisfactory, since $|LC|$ is
not related to the entanglement properties of the class. On the
other hand, Van den Nest, Dehaene, and De Moor have proposed a
finite set of invariants that characterizes all classes
\cite{VDD05}. However, this set has more than $2 \times 10^{36}$
invariants already for $n=7$. The problem of obtaining a minimum set
of invariants capable of distinguishing all classes with $n \le 8$
qubits will be addressed elsewhere \cite{CLMP09}.

Another weak point in the method is that the precise value of $E_S$
is still unknown for some classes. The good news is that, for most
of these classes, the value might be fixed if we knew the value for
the $5$-qubit ring cluster state, which is the first graph state in
the classification for which the value of $E_S$ is unknown
\cite{HEB04, HDERVB06}. Unfortunately, we have not made any progress
in calculating $E_S$ for the $5$-qubit ring cluster state.

Table~\ref{MainTable} shows that there are no $8$-qubit graph states
with rank indexes $RI_{p}=[\nu_{j}^{p}]_{j=p}^{1}$ with $\nu_{j}^{p}
\neq 0$ if $j=p$, and $\nu_{j}^{p} = 0$ if $j < p$, i.e., with
maximal rank with respect to all bipartite splits, i.e., such that
entanglement is symmetrically distributed between all parties. These
states are robust against disentanglement by a few measurements.
Neither there are $7$-qubit graph states with this property
\cite{HEB04, HDERVB06}. This makes more interesting the fact that
there is a single $5$-qubit and a single $6$-qubit graph state with
this property \cite{HEB04, HDERVB06}.


\section*{Acknowledgments}


The authors thank H. J. Briegel, O. G{\"u}hne, M. Grassl, M. Hein,
C. Santana, and M. Van den Nest, for their help. This work has
benefited from the use of the program {\em nauty} \cite{McKay90} for
computing automorphism groups of graphs. A.C., A.J.L., and P.M.
acknowledge support from projects No.~P06-FQM-02243,
No.~FIS2008-05596, and No.~PAI-FQM-0239. J.R.P. acknowledges support
from projects No.~P06-FQM-01649, No.~MTM2008-05866-C03-01, and
No.~PAI-FQM-0164.


\end{document}